\documentclass[conference]{IEEEtran}

\usepackage{mathrsfs}
\usepackage{amsfonts}
\usepackage{ifpdf}
\usepackage{cite}
\usepackage{stfloats}
\ifCLASSINFOpdf
\usepackage[pdftex]{graphicx}
\else \fi
\usepackage[cmex10]{amsmath}
\usepackage{array}
\usepackage{mdwmath}
\usepackage{mdwtab}
\usepackage{eqparbox}
\usepackage[tight,footnotesize]{subfigure}
\usepackage{algorithmic}
\usepackage{algorithm}
\usepackage{amsmath}
\usepackage{epsfig}
\usepackage{ae}
\usepackage{amssymb}
\usepackage{times}
\usepackage{amsmath}   % add1
\usepackage{booktabs}  % add2
\usepackage{threeparttable} %add3
\usepackage{multirow}       %add4
\usepackage{xcolor}
\usepackage{paralist}

\ifCLASSINFOpdf
\else
\fi
\begin{document}
\title{End-to-end Delay in Two Hop Relay MANETs with Limited Buffer}

\author{\IEEEauthorblockN{Jia~Liu}
\IEEEauthorblockA{School of Systems Information Science\\
Future University Hakodate\\
Hakodate, Hokkaido, Japan\\
jliu871219@gmail.com}
\and
\IEEEauthorblockN{Yang~Xu}
\IEEEauthorblockA{State Key Laboratory of ISN\\
Xidian University\\
Xi'an, China\\
yxu@xidian.edu.cn}
\and
\IEEEauthorblockN{Xiaohong~Jiang}
\IEEEauthorblockA{School of Systems Information Science\\
Future University Hakodate\\
Hakodate, Hokkaido, Japan\\
jiang@fun.ac.jp}
}
\maketitle

\begin{abstract}
Despite lots of literature has been dedicated to researching the delay performance in two-hop relay (2HR) mobile ad hoc networks (MANETs), however, they usually assume the buffer size of each node is infinite, so these studies are not applicable to and thus may not reflect the real delay performance of a practical MANET with limited buffer. To address this issue, in this paper we explore the packet end-to-end delay in a 2HR MANET, where each node is equipped with a bounded and shared relay-buffer for storing and forwarding packets of all other flows. The transmission range of each node can be adjusted and a group-based scheduling scheme is adopted to avoid interference between simultaneous transmissions, meanwhile a handshake mechanism is added to the 2HR routing algorithm to avoid packet loss. With the help of Markov Chain Theory and Queuing Theory, we develop a new framework to fully characterize the packet delivery processes, and obtain the relay-buffer blocking probability (RBP) under any given exogenous packet input rate. Based on the RBP, we can compute the packet queuing delay in its source node and delivery delay respectively, and further derive the end-to-end delay in such a MANET with limited buffer.  

\end{abstract}


\IEEEpeerreviewmaketitle

\section{Introduction}
A mobile ad hoc network (MANET) can be defined as a fully self-organizing system where mobile nodes freely communicate with each other without any infrastructure or centralized administration \cite{Perkins_BOOK01}. In such networks, the traditional routing algorithms like AODV \cite{Perkins_AODV99} and DSR \cite{Johnson_DSR96} can not adapt to the highly dynamic topology, while the routing algorithms based on opportunistic transmission like two-hop relay routing scheme \cite{Grossglauser_Tse_2001} which is first proposed by Grossglauser and Tse, and its variants \cite{Neely_IT05,Liu_TWC11,Liu_TON12,Liu_TC13,Liu_TWC12} are widely applied due to their simplicity and highly efficiency. Therefore, a critical issue of natural interest is how to thoroughly understand the performance of such networks \cite{Andrews_CM08,Goldsmith_CM11}. 

In our previous work \cite{Liu_PIMRC14}, we investigated the throughput and capacity of a buffer-limited MANET. So in this paper, we further extend the network model to a more general scenario and explore the end-to-end delay performance. By now, a lot of work has been done to analyze the packet delay in a class of 2HR MANETs. Neely and Modiano \cite{Neely_IT05} studied the end-to-end delay under several routing schemes such as 2HR with or without redundancy, 2HR with feedback and multi-hop relay. They developed a fundamental tradeoff between delay and throughput as $\mathbf{D}/\lambda=O(n)$. Groenevelt \emph{et al.} \cite{Groenevelt_PE05} also derived expressions for message delivery delay in closed-form. There also exist many scaling law results for the delay performance in MANETs under various mobility models, like under the random walk model in \cite{Gamal_IT06}, under the restricted mobility model in \cite{Mammen_IT07}, under Brownian motion model in \cite{Gamal_INFOCOM04,Lin_TON06}, and under hybrid random walk models in \cite{Sharma_TON07}. Recently, Liu \emph{et. al} explored the packet delay under $f$-cast relay algorithm \cite{Liu_TWC11}, generalized two-hop relay algorithm \cite{Liu_TON12}, and probing-based two-hop relay algorithm  \cite{Liu_TWC12}, respectively. 

However, it is notable that all these works mentioned above assumed the buffer size of each node is infinite to make their analysis tractable. Actually, this assumption never holds for a realistic MANET. Even in some scenarios, in order to save the networking cost, or due to the scarce resource in a terminal node (small size, low computing capability and so on), the buffer space equipped for each node is very limited. Thus, these studies are not applicable to and may not reflect the real delay performance of a practical MANET with limited buffer.

As a first step towards this end, this paper explores the packet end-to-end delay performance for a 2HR MANET, where each node is equipped with a limited relay-buffer, which is shared by all other traffic flows to temporarily store the forward their packets \cite{Le_TON12}. In order to avoid the interference between simultaneous transmissions, a group-based transmission scheduling scheme is adopted. While in order to avoid the packet loss when the relay-buffer of receiver is blocked, a handshake mechanism is added in the 2HR routing algorithm. The main contributions of this paper are summarized as follows.

\begin{itemize}
\item
A theoretical framework is developed to fully capture the packet arrival and departure processes in both source node and relay node, respectively. Based on this framework, we obtain the packet occupancy distribution in a relay buffer, and further derive the relay-buffer blocking probability (RBP) under any given exogenous input rate.

\item
The service rate of a source node can be computed by utilizing the RBP. Based this service rate and Queuing theory, we derive the queuing delay of a packet in its source node.  

\item 
With the help of RBP and the absorbing Markov Chain theory, we further derive the packet delivery delay. Finally, the packet end-to-end delay can be obtained by incorporating the queuing delay with the delivery delay. 
\end{itemize}

The remainder of this paper is organized as follows. Section~\ref{section:preliminaries} introduces the system models, transmission scheduling, routing algorithm and some basic definitions. Section~\ref{section:RBP} provides the theoretical framework to analyze the packet deliver processes and obtain the RBP. Based on the computation of RBP, the packet queuing delay and delivery delay are derived in Section~\ref{section:delay}. Finally, Section~\ref{section:numerical_results} provides the numerical results and Section~\ref{section:conclusion} concludes this paper.

%-------------------------------------------------------------------------new section------------------------------------------------------------------------%
\section{Preliminaries} \label{section:preliminaries}
This section introduces the system models, transmission scheduling, routing algorithm and some basic definitions involved in this paper.

\subsection{System Models} 

\begin{figure}[!t]
\centering
{
\subfigure[A cell-partitioned network with a general transmission range.]
{\includegraphics[width=1.6in]{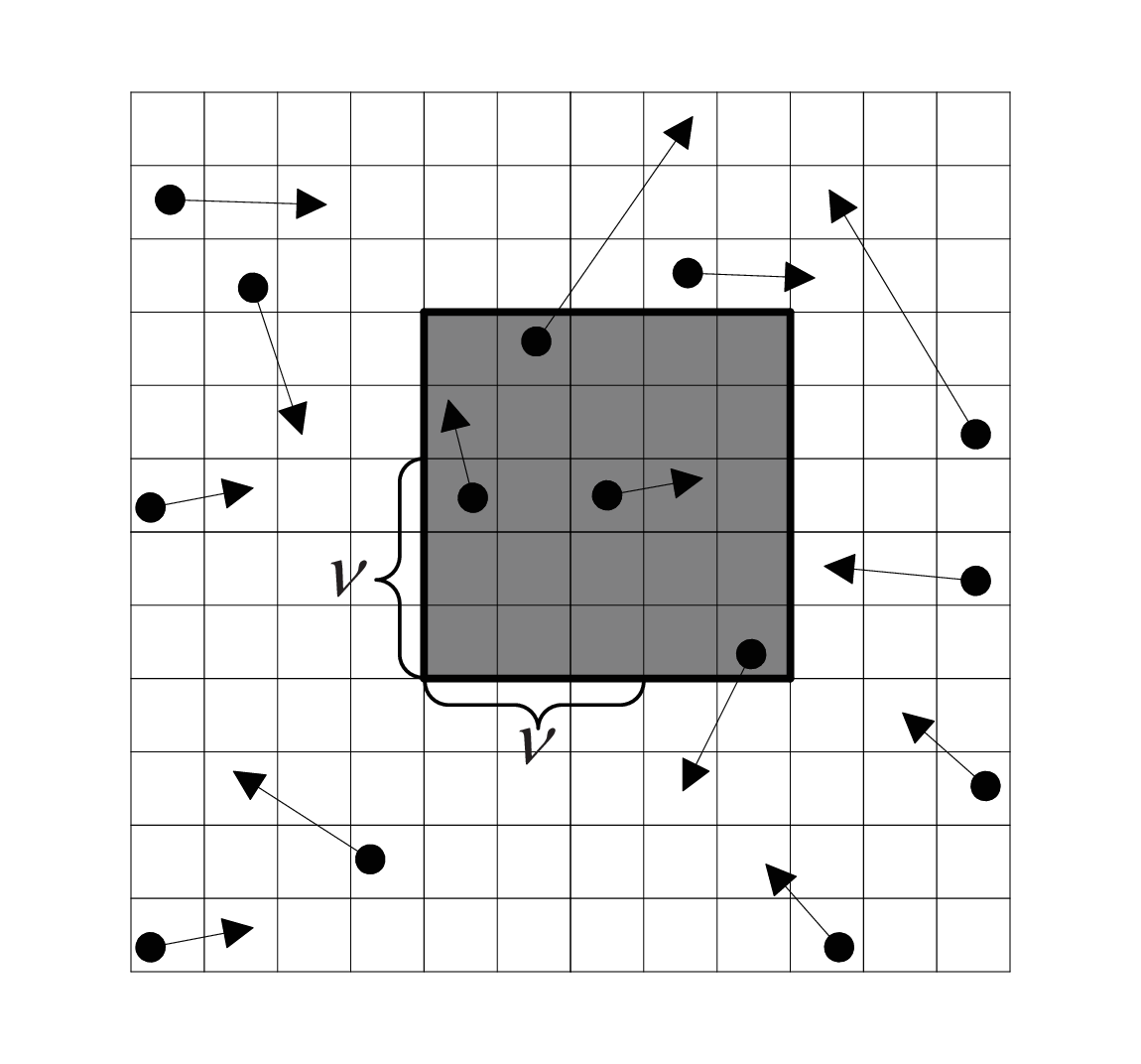} \label{fig:cell_partitioned} }
\hfill
\subfigure[Illustration of group-based scheduling.]
{\includegraphics[width=1.6in]{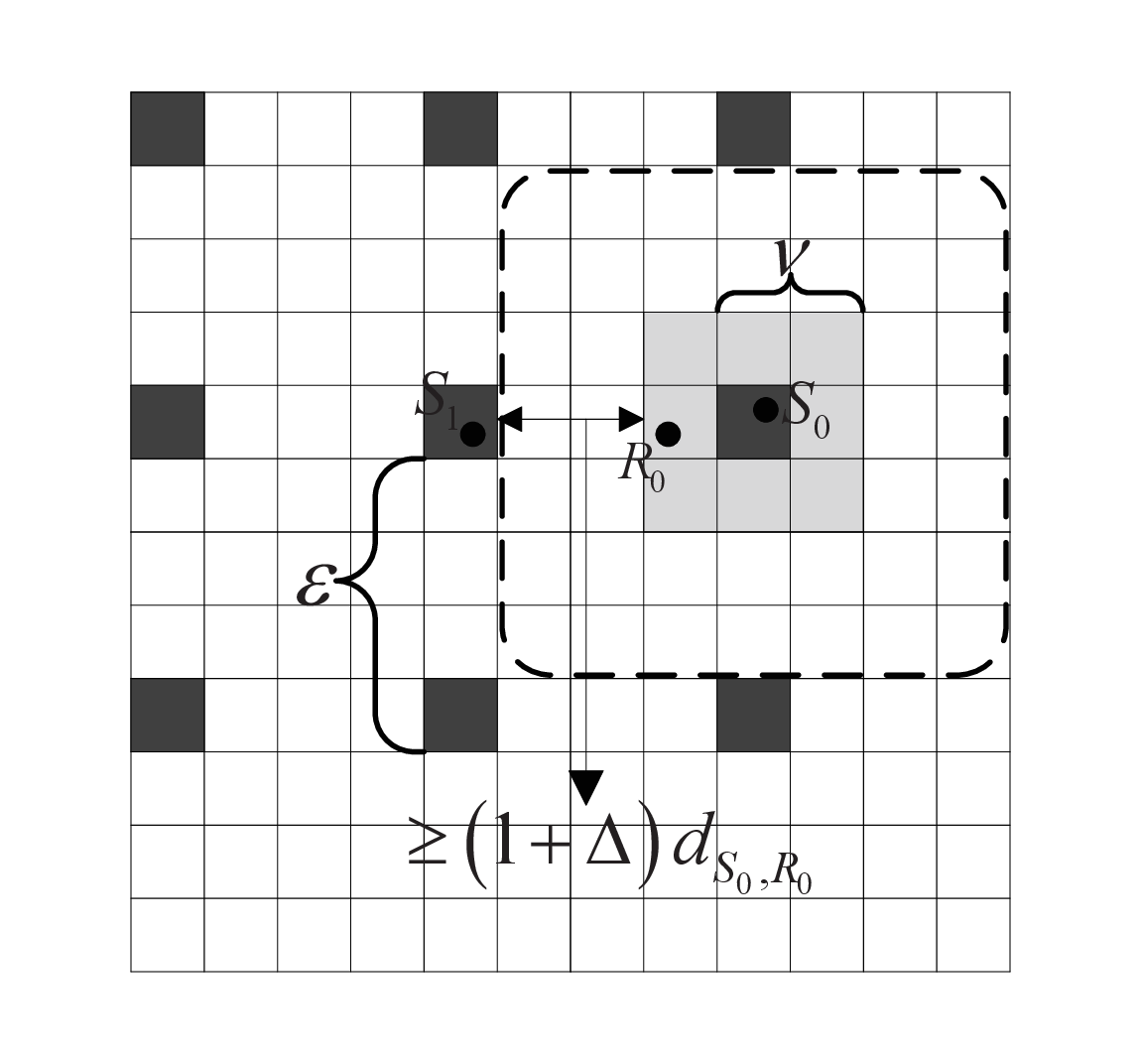} \label{fig:group-based_scheduling} }
}
\caption{Cell partitioned MANET and group-based scheduling.}
\end{figure}

\emph{Network model}: As previous works \cite{Neely_IT05,Liu_TON12}, we consider a time-slotted and cell-partitioned network model, where the network is partitioned into $m\times m$ nonoverlapping cells of equal size and $n$ mobile nodes roam from cell to cell according to the independent and identically distributed (i.i.d) mobility model \cite{Grossglauser_Tse_2001}. The time-slot has a fixed length and is uniformed to exact one packet transmission. The transmission range of each node is same and can cover a set of cells which have a horizontal and vertical distance of no more than $\nu -1$ cells away from its own cell, as illustrated in Fig.~\ref{fig:cell_partitioned}.

\emph{Traffic model}: The popular permutation traffic model \cite{Ciullo_TON11} is adopted. There are in total $n$ distinct unicast traffic flows, each node is the source of a traffic flow and meanwhile the destination of another traffic flow. Without loss of generality, as shown in \cite{Neely_IT05}, we assume $n$ is even and the source-destination pairs are composed as follows: $1 \leftrightarrow 2$, $3 \leftrightarrow 4$, $\cdots$, $(n-1) \leftrightarrow n$. The exogenous packet arrival at each node is a Bernoulli process with rate $\lambda$ packets/slot.

\emph{Interference model}: We adopt the famous protocol model \cite{Gupta_IT00} to account for the interference between simultaneous transmissions. By applying the protocol model, when node $i$ transmits packets to node $j$, this transmission is successful if and only if: 
\begin{itemize}
\item[1)]
Node $j$ is within the transmission range of node $i$.

\item[2)]
$d_{k,j} \geq (1+\Delta) d_{i,j}$, for any other concurrent transmitter $k$, where $d_{i,j}$ denotes the distance between $i$ and $j$, $\Delta$ is a guard factor determined by the protocol model.
\end{itemize}

\emph{Buffer Constraint}:
As the available study on buffer-limited wireless networks \cite{Le_TON12}, we consider a practical buffer constraint. Each node in the MANET has two queues, one local queue with unlimited buffer size for storing the self-generated packets, and one relay queue with fixed size $B$ for storing the packets coming from all other $n-2$ traffic flows. We adopt this buffer constraint here mainly due to the following reasons. First, in a practical network, each node usually reserves a much larger buffer space for storing its own packets rather than the relay packets. Second, even though the local buffer space is not enough when bursty traffic comes, the upper layer can execute congestion control to avoid the loss of local packets. Thus, our network model can be served as a well approximation for a realistic MANET. 

\subsection{Group-Based Transmission Scheduling}
As a inherent feature of wireless networks, the interference between simultaneous transmissions is a critical issue that should be carefully considered. We adopt here the group-based transmission scheduling which has been extensively applied in previous studies \cite{Liu_TON12,Ciullo_TON11}. As illustrated in Fig.~\ref{fig:group-based_scheduling}, all cells are divided into distinct groups, where any two cells in the same group have a horizontal and vertical distance of some multiple of $\epsilon$ cells. Thus, the MANET has $\epsilon^2$ groups and each group contains $K=\lfloor m^2 / \epsilon^2 \rfloor$ cells. Each group becomes active every $\epsilon^2$ time slots and each cell of an active group allows one node to conduct packet transmission. By applying our interference model, $\epsilon$ should be satisfied that
\[
(\epsilon - \nu) \cdot \frac{1}{m} \geq (1+\Delta) \sqrt{2} \nu \cdot \frac{1}{m}.  \nonumber
\]
On the other hand, in order to allow as many simultaneous transmissions as possible, $\epsilon$ is determined as
\begin{equation}
\epsilon=\min \{\lceil (1+\Delta) \sqrt{2} \nu+\nu \rceil,m \}. \label{eq:epsilon}
\end{equation}

\subsection{Handshake-Based Two Hop Relay Routing Algorithm}
Notice that in a buffer-limited MANET with 2HR for packet delivery, when a source node want to send a packet to a relay whose relay queue is full, then this transmission fails, and leads to packet loss and energy waste. To solve this problem, a handshake mechanism is introduced into the traditional 2HR algorithm, termed as H2HR. With H2HR, before each source-to-relay (s-r) transmission, the source node initiates a handshake with the relay node to confirm its relay-buffer occupancy state, once the relay queue is full, the source node cancels this transmission. At any time slot, for an active cell $c$,  it executes the H2HR algorithm as shown in Algorithm~\ref{algorithm:H2HR}. 

\begin{algorithm}[!ht]
\caption{H2HR algorithm}
\label{algorithm:H2HR}
\begin{algorithmic}[1]
\IF{There exist source-destination pairs, which one node of this pair is within $c$, another node is within the transmission range of $c$}
	\STATE With equal probability, randomly select such a pair to do source-to-destination (s-d) transmission.
\ELSIF{There exist some nodes in $c$, and some other nodes in the transmission range of $c$}
	\STATE With equal probability, randomly select one node in $c$ as the transmitter.
	\STATE With equal probability, randomly select another node within the transmission range of $c$ as the receiver.
  \STATE Flips an unbiased coin.
  \IF{It appears the head}
	  \STATE \textbf{The transmitter initiates a handshake with the receiver to check whether the relay queue is full.}
		\IF{ \textbf{The relay queue of receiver is not full}}
			\STATE \textbf{The transmitter conducts a s-r transmission.}
		\ELSE
			\STATE \textbf{The transmitter remains idle.}
		\ENDIF
	\ELSE
	  \STATE The transmitter conducts a relay-to-destination (r-d) transmission.
	\ENDIF
\ELSE
  \STATE $c$ remains idle.
\ENDIF

\end{algorithmic}
\end{algorithm}

\subsection{Basic Definitions}
\textbf{Relay-buffer Blocking Probability (RBP)}: For the concerned MANET with a given exogenous packet arrival rate $\lambda$ to each node, the relay-buffer blocking probability $p_b(\lambda)$ of a node is defined as the probability that the relay queue of this node is full.

\textbf{Queuing Delay}: The queuing delay of a packet is defined as the interval between the time this packet arrives at its source node and the time it takes to arrive at the head of local queue.

\textbf{Delivery Delay}: The delivery delay of a packet is defined as the interval between the time this packet arrives at the head of its local queue and the time it takes to be delivered to the destination node.

\textbf{End-to-end Delay}: The end-to-end delay of a packet is defined as the interval between the time this packet arrives at its source node and the time it takes to be delivered to its destination node. Obviously, the end-to-end delay of a packet is the sum of its queuing delay and delivery delay.

%-------------------------------------------------------------------------new section------------------------------------------------------------------------%
\section{Analysis for Packet Delivery Processes} \label{section:RBP}
In this section, we present the theoretical framework which help us fully characterize the complicated packet delivery processes and further compute the PBP.

\subsection{Some Basic Probabilities}
Considering a given time slot and a given active cell $c$, we denote by $p$ the probability that there are at least one node within $c$ and another node within the transmission range of $c$, and denote by $q$ the probability that there are at least one source-destination pair, one node of this pair is within $c$ and another one is within the transmission range of $c$. Based on the results of \cite{Gao_2013}, $p$ and $q$ are determined as
\begin{align}
& p=\frac{1}{m^{2n}}[m^{2n}-(m^2-1)^n-n(m^2-l)^{n-1}], \label{eq:p} \\
& q=\frac{1}{m^{2n}}[m^{2n}-(m^4-2l+1)^{n/2}], \label{eq:q}
\end{align} 
where $l=(2\nu -1)^2$. We denote by $p_{sd}$, $p_{sr}$ and $p_{rd}$ the probabilities that in a time slot a node obtains the opportunity to conduct s-d, s-r and r-d transmission, respectively. Similar to \cite{Liu_PIMRC14}, we have 
\begin{align}
& p_{sd}=\frac{K}{n}q, \label{eq:p_sd} \\
& p_{sr}=p_{rd}=\frac{K}{2n}(p-q). \label{eq:p_sr}
\end{align} 

\subsection{Delivery Processes in Local Queue and Relay Queue}

\begin{figure}[!t]
\centering\includegraphics[width=3.0in]{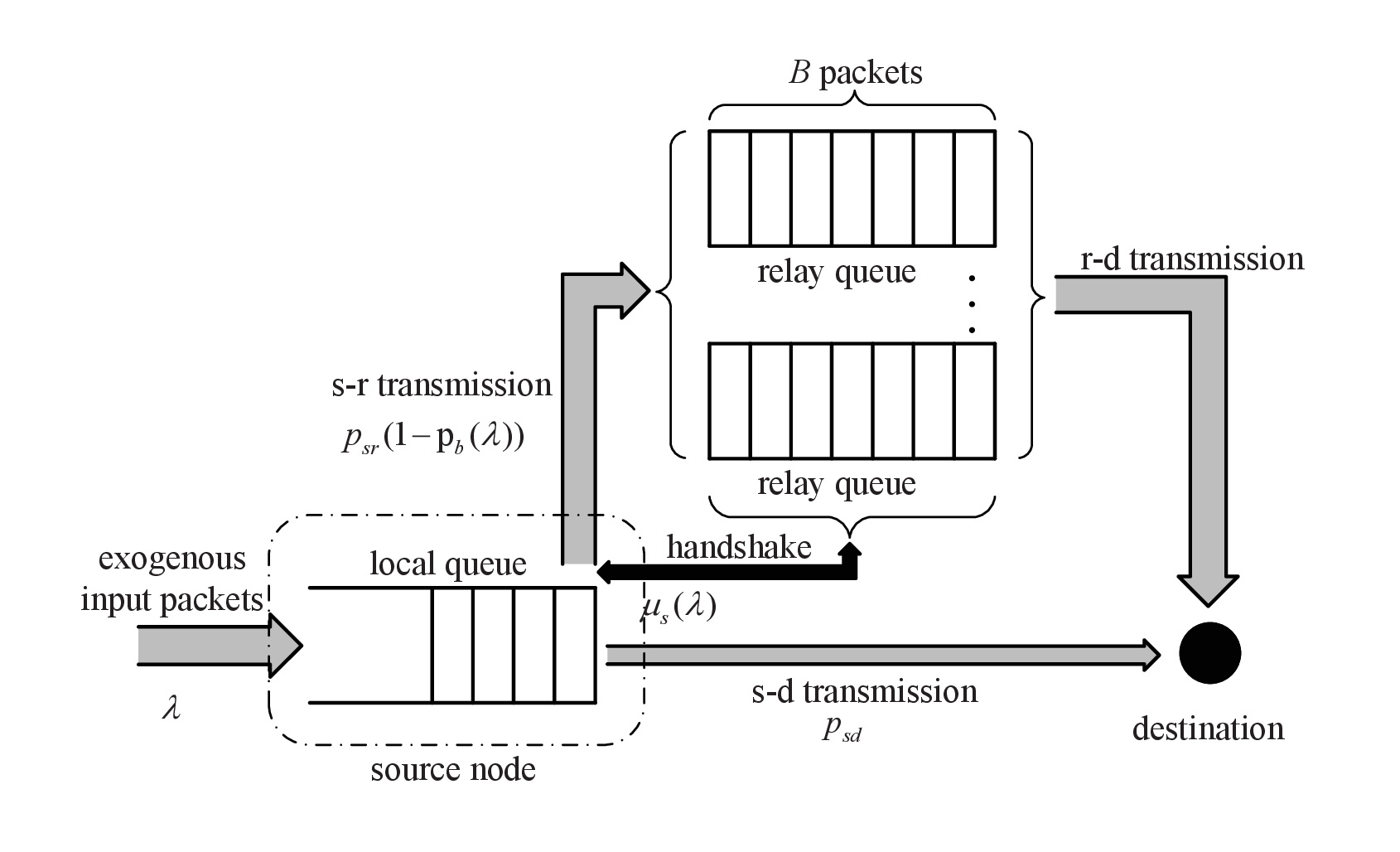}\caption{Illustration for packet delivery processes under H2HR algorithm.}
\label{fig:H2HR}
\end{figure}

The packet delivery processes under H2HR algorithm is illustrated in Fig.~\ref{fig:H2HR}. The local queue can be represented as a Bernoulli/Bernoulli queue, where in every time slot a new packet will arrive with probability $\lambda$, and a corresponding service rate $\mu_s(\lambda) $ which is determined as
\begin{equation}
\mu_s(\lambda)=p_{sd}+p_{sr} \cdot (1-p_b(\lambda)). \label{eq:mu_s}
\end{equation}
Due to the reversibility of Bernoulli/Bernoulli queue \cite{Daduna_BOOK01}, its output process is also a Bernoulli flow with rate $\lambda$.

As shown in Fig.~\ref{fig:H2HR}, the ratio of packets transmitted to a relay node is $\frac{p_{sr}(1-p_b(\lambda))}{\mu_s(\lambda)}$. Due to the i.i.d mobility model, each of the $n-2$ relay nodes will receive this packet with equal probability. On the other hand, for a specific node, the packets from all other $n-2$ flows will arrive its relay queue. Then the packet arrival rate at a relay queue $\lambda_r$ can be determined as
\begin{eqnarray}
&\lambda_r \cdot (1-p_b(\lambda))+ 0 \cdot p_b(\lambda)=  \frac{(n-2)\lambda \cdot p_{sr}\left(1-p_b(\lambda)\right)}{\mu_{s}(\lambda) (n-2)}, &\nonumber \\ 
&\Rightarrow \lambda_r= \frac{\lambda p_{sr}}{\mu_s(\lambda)}. &
\label{eq:lambda_r}
\end{eqnarray}

We denote by $\mu_r(k)$ that the service rate of relay queue when it contains $k$ packets, $0 \leq i \leq B$. According to the results in \cite{Liu_PIMRC14}, we have
\begin{align}
\mu_r(k) &=\sum_{i=1}^{k} \left\{ \frac{ \binom{n-2}{i} \cdot \binom{k-1}{i-1}}
{\binom{n-3+k}{k}}\cdot\frac{i p_{rd}}{n-2}\right\} \nonumber \\
&=\frac{k}{n-3+k} \cdot p_{rd}. \label{eq:mu_r}
\end{align} 

Since the relay queue cannot forward and receive a packet at the same time slot, then it can be modeled as a discrete Markov chain. We use $\mathbf{\Pi}=(\pi_0,\pi_1,\cdots,\pi_B)$ to denote the limit occupancy distribution on relay queue, then we have \cite{Liu_PIMRC14}
\begin{align}
\pi_0&=\frac{1}{\sum_{i=0}^B{\mathrm{C}_i \cdot \rho_s(\lambda)^i}}, \label{eq:pi_0} \\
\pi_k&=\frac{\mathrm{C}_k \cdot \rho_s(\lambda)^k} {\sum_{i=0}^B{\mathrm{C}_i  \cdot \rho_s(\lambda)^i}},  \quad 0<k\leq B  \label{eq:pi_k}
\end{align}
where $\mathrm{C}_i=\binom{n-3+i}{i}$ and $\rho_s(\lambda)=\frac{\lambda}{\mu_s(\lambda)}$. 

When a relay queue contains $B$ packets, this queue is full. Thus we have 
\begin{equation}
p_b(\lambda)=\pi_B=\frac{\mathrm{C}_B \cdot \rho_s(\lambda)^B} {\sum_{i=0}^B{\mathrm{C}_i  \cdot \rho_s(\lambda)^i}} \label{eq:p_b}
\end{equation}

Notice that given a exogenous input rate $\lambda$, equation (\ref{eq:p_b}) contains only one unknown quantity $p_b(\lambda)$. By solving equation (\ref{eq:p_b}), we can then obtain the RBP $p_b(\lambda)$ under any exogenous input rate $\lambda$.

%-------------------------------------------------------------------------new section------------------------------------------------------------------------%
\section{Delay Performance} \label{section:delay}
With the help of RBP, in this section we further analyze the packet delay performance in a buffer-limited MANET. We denote by $\mathbf{D}$, $\mathbf{W}$ and $\mathbf{T}$ the packet end-to-end delay, queuing delay and delivery delay, respectively. Then we have $\mathbf{D}=\mathbf{W}+\mathbf{T}$.

\subsection{Queuing Delay}
Given the exogenous input rate $\lambda$, the RBP $p_b(\lambda)$ can be obtained by equation (\ref{eq:p_b}), further the service rate of local queue $\mu_s(\lambda)$ (in the rest of this paper, $p_b(\lambda)$ and $\mu_s(\lambda)$ are abbreviated as $p_b$ and $\mu_s$ if there is no ambiguous) can be determined by formula (\ref{eq:mu_s}). Then, the average queue length of the local queue (Bernoulli/Bernoulli queue) $\mathbb{E}\{L_s\}$ is given by \cite{Daduna_BOOK01}
\begin{equation}
\mathbb{E}\{L_s\}=\frac{\lambda-\lambda^{2}}{\mu_s-\lambda}. \label{eq:L_s}
\end{equation}  
According to the Little's Theorem, the average delay of a packet in its local queue $\mathbb{E}\{D_s\}$ is 
\begin{equation}
\mathbb{E}\{D_s\}=\frac{1-\lambda}{\mu_s-\lambda}. \label{eq:D_s}
\end{equation} 
Then, the queuing delay $\mathbf{W}$ is determined as
\begin{equation}
\mathbb{E}\{\mathbf{W}\}=\mathbb{E}\{D_s\}-\frac{1}{\mu_s}=\frac{\lambda(1-\mu_s)}{\mu_s(\mu_s-\lambda)} \label{eq:W}.
\end{equation}  

\subsection{Delivery Delay}
\begin{figure}[!t]
\centering
\includegraphics[width=2.0in]{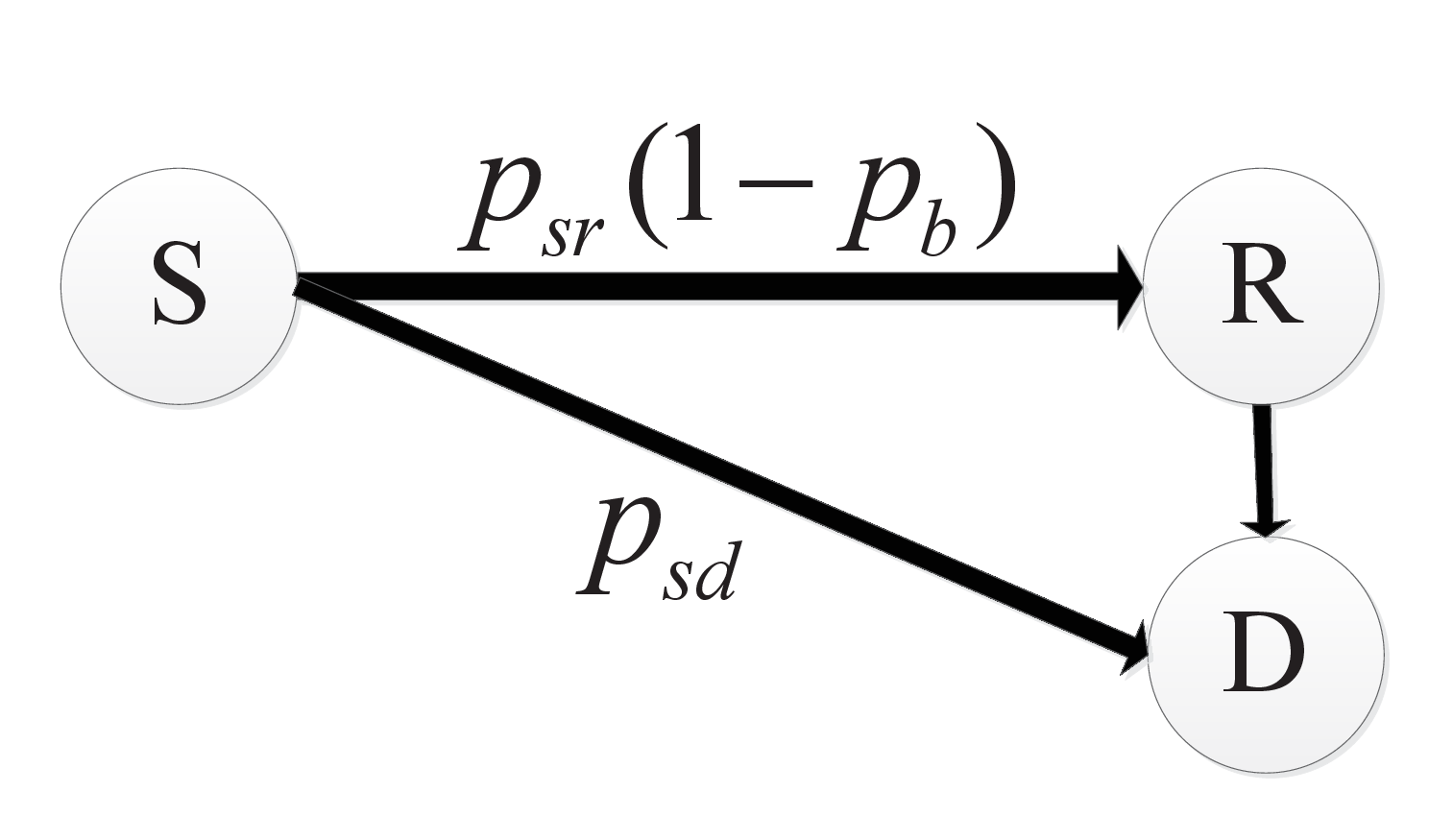}\caption{The absorbing Markov chain for a focused packet delivery.}
\label{fig:absorbing_Markov}
\end{figure}

Without loss of generality, we focus on a packet $y$ which is in the head of a local queue. As illustrated in Fig.~\ref{fig:absorbing_Markov}, in the next time slot, packet $y$ will be transmitted to its destination node with probability $p_{sd}$, to a relay node with probability $p_{sr}\cdot(1-p_b)$, and stays in the local queue with probability $1-\mu_s$, which forms an Absorbing Markov Chain. We denote by $X_{\mathbf{S}}$ and $X_{\mathbf{R}}$ the average transition time from the transient states $\mathbf{S}$ and $\mathbf{R}$ to the absorbing state $\mathbf{D}$, respectively. Then we have
\begin{equation}
X_{\mathbf{S}}=1+X_{\mathbf{S}} \cdot (1-\mu_s) + X_{\mathbf{R}} \cdot p_{sr} (1-p_b), 
\end{equation}
and
\begin{equation} 
\mathbb{E}\{\mathbf{T}\}=X_{\mathbf{S}}= \frac{1+X_{\mathbf{R}} \cdot p_{sr}(1-p_b)}{\mu_s}. \label{eq:E_T}
\end{equation}

We denote by $\mathbf{P}=(p_0,p_1,\cdots,p_{B-1})$ the probability that there are $i$ packets destined for the same node as $y$ is in front of $y$, when $y$ is transmitted into a relay queue. Notice that in a time slot, a node executes the r-d transmission with probability $p_{rd}$ which is shared by all the $n-2$ traffic flows equally, then we have
\begin{align}
X_{\mathbf{R}} &= p_0 \cdot \frac{n-2}{p_{sd}} +  2 p_1 \cdot \frac{n-2}{p_{sd}} + \cdots  B p_{B-1} \cdot \frac{n-2}{p_{sd}} \nonumber \\
&=  \frac{n-2}{p_{sd}} (p_0+2p_1+\cdots+Bp_{B-1}) \nonumber \\
&=  \frac{n-2}{p_{sd}} (1+p_1+2p_2 + \cdots+(B-1)p_{B-1}) \nonumber \\
& =  \frac{n-2}{p_{sd}} (1+ \mathbb{E} \{ L_{r|nf}^{(1)} \}),  \label{eq:X_R}
\end{align}
where $\mathbb{E} \{ L_{r|nf}^{(1)} \}$ denotes the average number of packets in a relay queue which are destined for a same node, under the condition that this relay is not full.

We denote by $\mathbf{\Pi'}=(\pi_0',\pi_1',\cdots,\pi_{B-1}')$ the occupancy distribution on relay queue given that this relay queue is not full. Then we have
\begin{equation}
\pi_k'=\frac{\pi_k}{1-\pi_B}=\frac{\mathrm{C}_k \rho_s^k}{\sum_{i=0}^{B-1}{\mathrm{C}_i  \cdot \rho_s^i}}. \label{eq:pi_k'}
\end{equation}
Thus, the average queue length of a relay queue given that it is not full $\mathbb{E}\{ L_{r|nf} \}$ is determined as
\begin{equation}
\mathbb{E}\{ L_{r|nf} \}= \sum_{k=0}^{B-1}{i\cdot \pi_k'} = \frac{\sum_{i=0}^{B-1}{ i \mathrm{C}_i  \cdot \rho_s^i}}{\sum_{i=0}^{B-1}{\mathrm{C}_i  \cdot \rho_s^i}}.
\label{eq:L_r|nf}
\end{equation}
Then, $\mathbb{E} \{ L_{r|nf}^{(1)} \}$ is determined as
\begin{equation}
\mathbb{E} \{ L_{r|nf}^{(1)} \} = \frac{\mathbb{E}\{ L_{r|nf} \}}{n-2} \label{eq:L_r^1}.
\end{equation}
Substituting the results of (\ref{eq:X_R}), (\ref{eq:L_r|nf}) and (\ref{eq:L_r^1}) into (\ref{eq:E_T}), the average packet deliver delay $\mathbb{E}\{\mathbf{T}\}$ is further determined. 

Finally, the expectation of packet end-to-end delay in the concerned buffer-limited MANET is determined as
\begin{equation}
\mathbb{E}\{\mathbf{D}\}=\mathbb{E}\{\mathbf{W}\}+\mathbb{E}\{\mathbf{T}\}
\end{equation}
 
%-------------------------------------------------------------------------new section------------------------------------------------------------------------%
\section{Numerical Results} \label{section:numerical_results}
\begin{figure}[!t]
\centering
\includegraphics[width=3.0in]{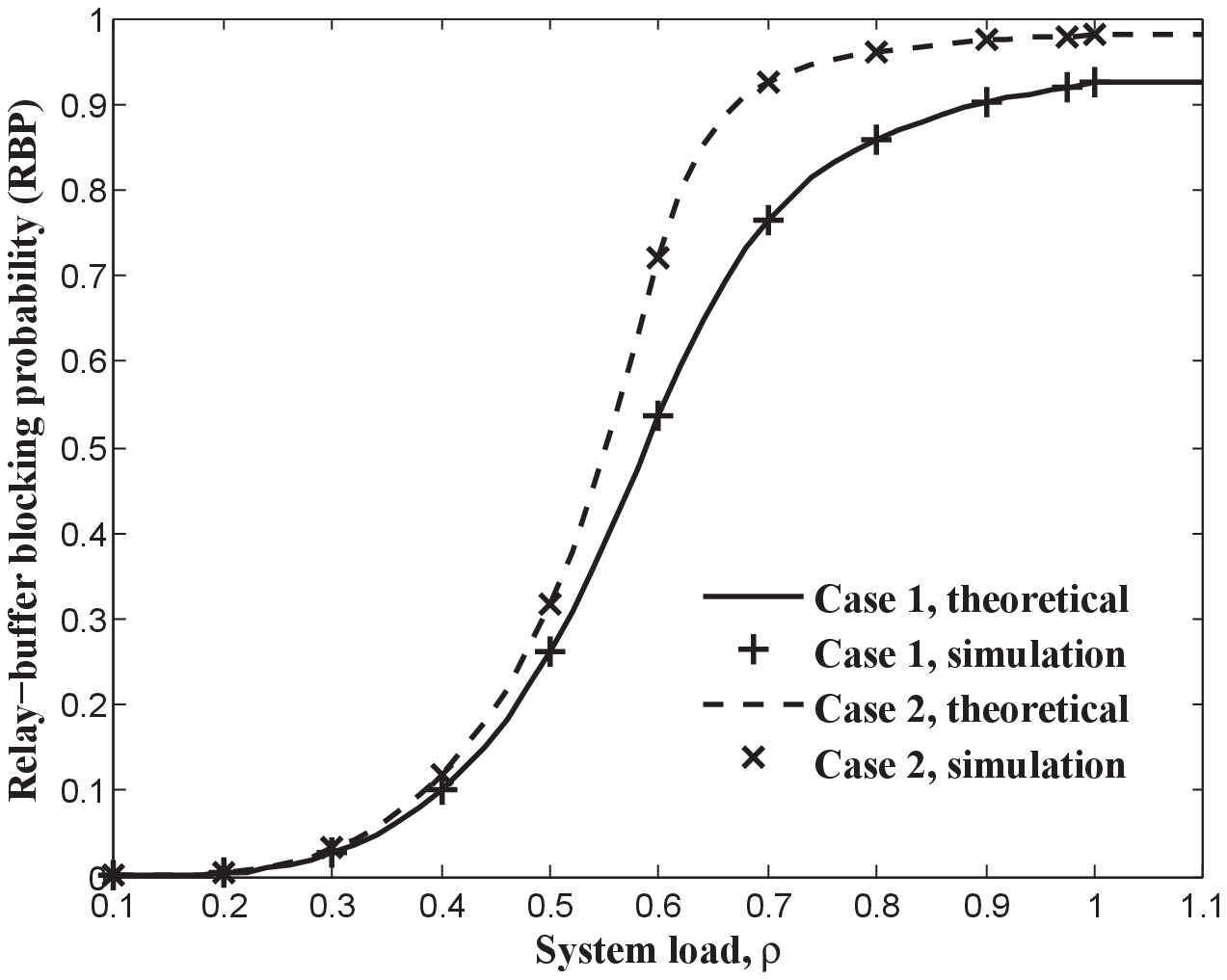} \caption{Relay-buffer probability RBP vs. system load $\rho$}
\label{fig:RBP}
\end{figure}

We conduct a C++ simulator to simulate the behaviors of MANETs considered in this paper. In our simulations, we set $\nu=1$, $\Delta=1$ \cite{ns-2} and choose two network scenarios of (case 1: $n=100, m=8, B=8$) and (case 2: $n=400, m=16, B = 8$). The theoretical RBP results are computed by the equation (\ref{eq:p_b}). While, to obtain the simulated RBP results, we focus on a specific node and count the number of time slots that its relay-buffer is full over a period of $2\times10^8$ time slots, and then calculate the ratio. Fig.~\ref{fig:RBP} compares the theoretical curves with the simulated results under a variable system load $\rho$, where $\rho=\frac{\lambda}{\mu_s(\lambda_0)}$, $\mu_s(\lambda_0)$ satisfies that $\mu_s(\lambda_0)=\lambda_0$ and thus is the maximal throughput the MANET can support \cite{Liu_PIMRC14}. We can see that for both the two cases, the simulated RBP can match the theoretical curves nicely, indicating that our theoretical framework is highly efficient to capture the packet delivery processes in a buffer-limited MANET with H2HR algorithm.

\begin{figure}[!t]
\centering
\includegraphics[width=3.0in]{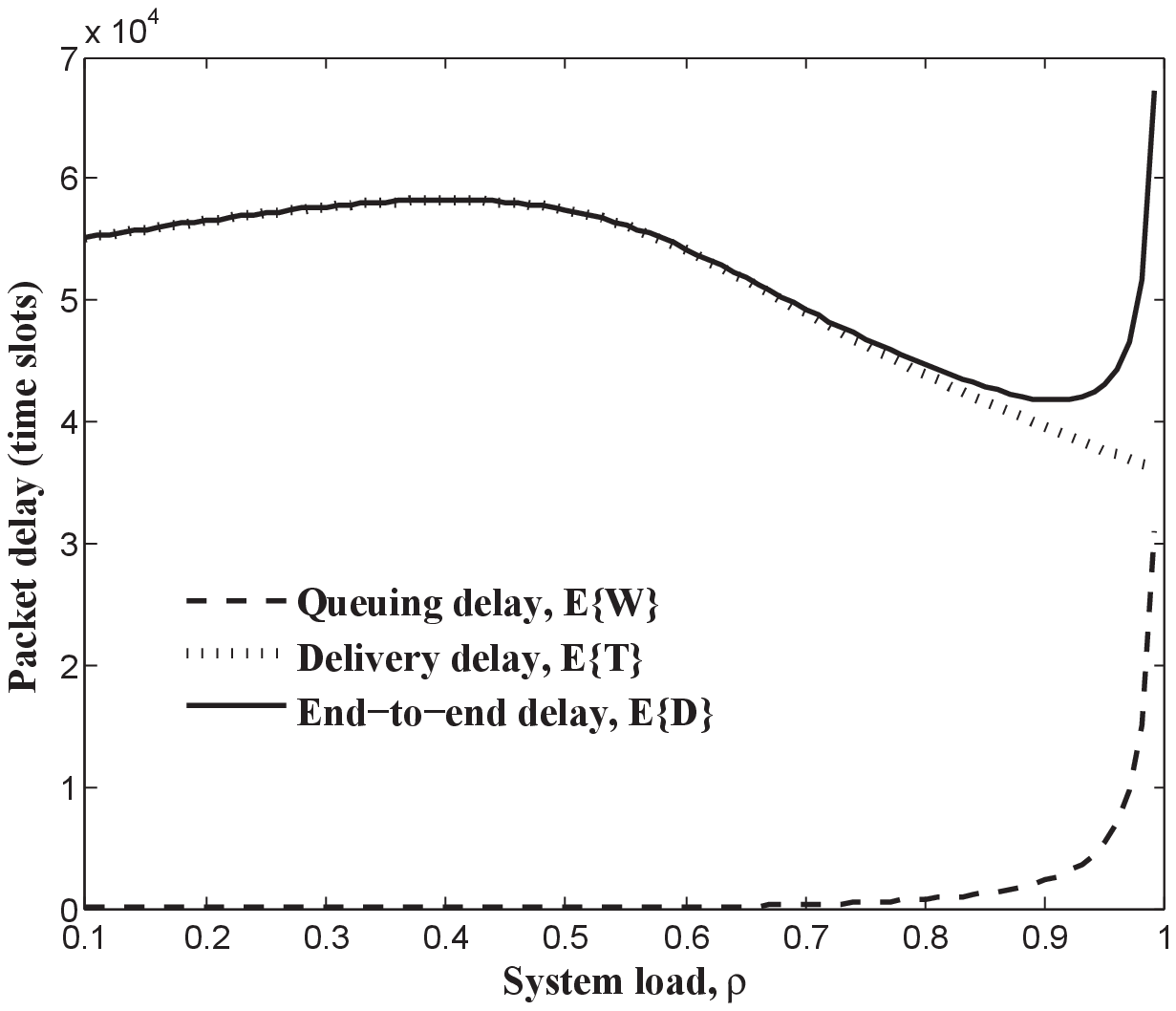} \caption{Packet delay vs. system load $\rho$}
\label{fig:delay_rho}
\end{figure}

Based on the RBP, we then show the packet delay performance with the system load $\rho$ under the network setting of ($n=50, m=5, B=8, \nu=1, \Delta=1$). The results of packet queuing delay, delivery delay and end-to-end delay are summarized in Fig.~\ref{fig:delay_rho}. We can see that when $\rho$ is small, the packet queuing delay $\mathbb{E}\{\mathbf{W}\}$ is small; as $\rho$ increases, $\mathbb{E}\{\mathbf{W}\}$ monotonically increases; when $\rho$ approaches $1$, $\mathbb{E}\{\mathbf{W}\}$ tends to infinity leading that the packet end-to-end delay is infinite. While, the packet delivery delay performance under the limited-buffer scenario is interesting, which increases first, then decreases. This is mainly due to the reason that the effects of the exogenous input rate on delivery delay are two folds. On one hand, a larger $\rho$ will lead to a longer relay queue length which further leads to a larger delay in a relay queue;
on the other hand, a larger $\rho$ will lead to a higher RBP, which means a lower ratio of packets conducted by s-r transmission, packets in the head of local queue are more likely to wait a direct s-d transmission opportunity, thus the delivery delay decreases.

%-------------------------------------------------------------------------new section------------------------------------------------------------------------%
\section{Conclusion} \label{section:conclusion}
In this paper, we focus on the packet delay performance of a MANET under finite buffer scenario. A group-based transmission scheduling is adopted for channel access, while a handshake-based two hop relay algorithm is adopted for packet delivery. For the concerned MANET, a theoretical framework has been developed to fully characterize the queuing processes of a packet and obtain the relay-buffer blocking probability. Based on this, we has derived the packet queuing delay and delivery delay, respectively. The results show that the packet end-to-end delay performance curve first rises and then declines as the exogenous rate grows, finally rises again and tends to infinity as the exogenous rate approaches the network throughput capacity.

\end{document}